\documentclass[12pt]{article}
\usepackage{eucal,  amssymb,amsmath,graphicx, amsfonts, latexsym}
\usepackage[T1]{fontenc}

 \usepackage{color}
 \usepackage{geometry}
\usepackage[applemac]{inputenc}
\usepackage{tikz}
\usetikzlibrary{decorations.pathreplacing}
\usetikzlibrary{calc}

\renewcommand{\thefootnote}{\arabic{footnote}}

\begin{document} \parskip=5pt plus1pt minus1pt \parindent=0pt
\renewcommand{\thefootnote}{\arabic{footnote}}
\title{Estimation in emerging epidemics: biases and remedies}
\author{Tom Britton\footnote{Dept. of Mathematics, Stockholm University, 10691 Stockholm, Sweden, <tomb@math.su.se>.} and Gianpaolo Scalia Tomba\footnote{Dept. of Mathematics, Univ. of Rome Tor Vergata, 00133 Rome, Italy, <scaliato@mat.uniroma2.it>.} \footnote{Corresponding author}}
\date{\today}
\maketitle

\begin{abstract}
When analysing new emerging infectious disease outbreaks one typically has observational data over a limited period of time and several parameters to estimate, such as growth rate, $R_0$, serial or generation interval distribution, latent and incubation times or case fatality rates. Also parameters describing the temporal relations between appearance of symptoms, notification, death and recovery/discharge will be of interest.
These parameters form the basis for predicting the future outbreak, planning preventive measures  and monitoring the progress of the disease. We study the problem of making inference during the emerging phase of an outbreak and point out potential sources of bias related to contact tracing, replacing generation times by serial intervals, multiple potential infectors or truncation effects amplified by exponential growth. These biases directly affect the estimation of e.g. the generation time distribution  and the case fatality rate, but can then propagate to other estimates, e.g. of $R_0$ and growth rate. Many of the traditionally used estimation methods in disease epidemiology may suffer from these biases when applied to the emerging disease outbreak situation.
We show how to avoid these biases based on proper statistical modelling. We illustrate the theory by numerical examples and simulations based on the recent 2014-15 Ebola outbreak to quantify possible estimation biases, which may be up to 20\% underestimation of $R_0$, if the epidemic growth rate is fitted to observed data or, conversely, up to 62\% overestimation of the growth rate if the correct $R_0$ is used in conjunction with the Euler-Lotka equation.
\end{abstract}

\section*{Significance statement}
In the early phase of an emerging disease that threatens to become epidemic or pandemic, it is important to quickly assess growth and disease related parameters in order to plan and monitor the progress of countermeasures. We describe some important sources of bias in such estimates and ways to reduce or eliminate them.

\section{Introduction}\label{sec-intro}

During the last decades, several new disease outbreaks have struck the human or domestified animal populations, e.g.\ SARS, foot and mouth disease, H1N1 influenza, and, more recently, Ebola. These outbreaks have in common the need for estimation of key parameters to be performed early on in the outbreak, in order to plan interventions and monitor the progress of the disease. Thus estimation must be performed in the  \emph{emerging} phase of an outbreak, when the \emph{number} of infected individuals is in the hundreds or at most thousands, while  the community \emph{fraction} of infected is still small. Typically the early numbers grow exponentially, as also predicted by mathematical epidemic models (e.g.\ Diekmann et al.\ (2013)).

There may be many complicating or limiting factors related to completeness of data, lack of detailed knowledge about the disease and other issues when analysing data from the early phase of the outbreak. Despite these complicating factors, the conclusions drawn from early analyses are usually highly valuable. The aim of the present paper is to identify and highlight some of the potential biases in the statistical analysis of emerging outbreaks inherent in the early phase itself and to illustrate how they can be propagated to parameter estimates and predictions. A further aim is  to give some fairly simple suggestions for how to reduce, or even remove, such biasing effects. 

The typical available data consist of reported numbers of confirmed cases per day or week, some case histories illustrating the course of the disease and some contact tracing data containing information about possible durations between onset of symptoms of infected individuals and their infectors, whereas little information is usually available about uninfected individuals and their amount of exposure. The epidemic models used in the statistical analyses are often of simple form, neglecting various heterogeneities. The use of simple models in these situations is motivated by the lack of detailed information but has  also recently been studied by Trapman et al.\ (2016) who show that neglecting population structures when making inference in emerging outbreaks has little effect. However, estimation in simple models can still be quite complicated. The complications are mainly due to three factors: 1) important events, such as times of infection, are usually unobserved, but instead some proxy measures such as onset of symptoms are available, 2) estimation of parameters of the epidemic process is based on observations up to some fixed time, implying that events occuring later are censored, and 3) the population of infectives is increasing (exponentially) with time.

In our investigation, we first discuss the effect of estimating the generation time distribution from observations of generation times observed backwards in time using contact tracing, i.e.\ the time between the infection time of an individual (the infectee) and that of his/her infector (rather than the infection time of the individuals he/she infects). The second problem we study is the effect of 
replacing generation times (the time between infections of an infector and an infectee) with the more commonly observed serial intervals (the time between onset of symptoms of an infector and an infectee). A third problem we discuss is how to treat the common situation, when contact tracing, where there is more than one potential infector of some of the cases, with the implication that the backward generation time or incubation time (time from infection to symptoms) is one out of several possible values. As it turns out, the overall biasing effect, if these problems are not considered, can be highly significant when estimating e.g.\ the basic reproduction number.  We also address the problem of estimating the case fatality rate of the disease. We then quantify the various biases that can arise in a realistic parameter setting, using estimates and assumptions from the recent Ebola outbreak in West Africa  (Team WER et al., 2014). It turns out that e.g.\ $R_0$ could be underestimated by as much as 20\%, backward observation of generation times and the treatment of the multiple possible infectors problem being the main potential sources of error in that parameter setting, but it should be noted that results could be even worse in other settings.

Below we first introduce the underlying stochastic epidemic model. Then, in Sections \ref{sec-backwards}   to \ref{sec-multiple}, we investigate how the three potential biases appear and how to reduce/remove their effects. In Section  \ref{sec-fatality}  we describe how the estimated case fatality may also be equipped with a bias during the emerging phase of an outbreak, and how this bias may be reduced. In Section \ref{sec-illustration}  we illustrate our findings with parameters inspired by the recent West-Africa Ebola outbreak, with the aim of quantifying how big  biases due to the various causes may be and report some interesting simulation results. Section \ref{sec-disc} is a brief discussion and, finally, some mathematical and numerical details are collected in Supplementary Information.

\section{The underlying model and some key epidemiological quantities}\label{sec-model}

We start by presenting the basic underlying epidemic model. We assume that individuals are at first susceptible and later may become infected, and that infected individuals may then infect other individuals over time starting from the time of  infection. The infection ends with death or recovery and subsequent immunity. The population is assumed to be a homogeneously mixing community of homogeneous individuals. Usually this is not the case, but including all potential heterogeneous aspects is typically not possible due to lack of data and time pressure to obtain results; besides, it has been shown by Trapman et al.\ (2016) that neglecting heterogeneity when analysing an emerging outbreak has little effect on estimates of fundamental parameters like $R_0$, and that the (small) effect is nearly always in the direction of being conservative. Since we model the initial phase of the outbreak, the depletion of susceptibles is considered as negligible. Also, we assume that individuals do not change their behaviour over the considered time period  as a consequence of the ongoing outbreak, nor are there yet any control measures put in place by health authorities or similar. Finally, we assume that there are no seasonal changes in transmission. Similar assumptions are often made in early estimation of emerging outbreaks (e.g.\ Team WER et al.\ (2014)). Predictions are made assuming that the disease spreading mechanism continues unaltered, reflecting what would presumably happen in the absence of control measures (these predictions are then compared with predictions including various preventive measures).

Traditionally, the population effects of such an infection have been modelled using compartmental models with separate states for e.g.\  susceptible, latent, infectious or recovered/removed individuals (SI, SEI, SIR and SEIR models, see e.g.\ Anderson \& May (1991) or Diekmann et al.\ (2013)). Recently, modelling has reverted to something akin to the original Kermack-McKendrick (1927) formulation, with emphasis on one single quantity, $\beta (s)$,  the average rate at which an infected individual infects new individuals $s$ time units after his/her time of infection, denoted the infection rate (or infectivity function, e.g.\ Diekmann et al.\ (2013)). The assumption of a homogeneous community implies that $\beta (s)$ is the same for all individuals, and the assumptions of no depletion of susceptibles, no preventive measures and no seasonal effects imply that $\beta (s)$ is independent of the time of infection of the individual. The previously mentioned compartmental models can all be translated to this framework. While the original treatment of the Kermack-McKendrick model was deterministic (Volterra type integral equations), statistical modelling requires a stochastic formulation which, in this case, corresponds to Crump-Mode-Jagers branching processes (see e.g.\ Jagers (1975)) in the initial phase of spread. It should be noted that the infectivity functions in a stochastic model may be different from individual to individual, although the average behaviour is the same, and that different stochastic models may have the same average behaviour (see e.g.\  Svensson (2015)).

The average infection rate $\beta (s)$ completely determines the basic reproduction number $R_0$ and the epidemic exponential growth rate $r$, as is well-known in epidemic modelling (see e.g.\ Diekmann et al.\ (2013)) and branching process theory (see e.g.\ Jagers (1975)). 

The mean number of infections during the infectious period, better known as the basic reproduction number and denoted $R_0$, is given by:
\begin{equation}
R_0=\int_0^\infty \beta (s)ds. \label{R_0}
\end{equation}
It is well-known that an epidemic can take off if and only if $R_0>1$, which we from now on assume.

Another important quantity is the so-called \emph{generation time} distribution $f_G(s)$, which is simply the infection rate scaled to make it a probability distribution:
\begin{equation}
f_G(s)=\frac{\beta (s)}{R_0}=\frac{\beta (s)}{\int_0^\infty \beta (u)du}. \label{gen-time}
\end{equation}
The generation time distribution is the probability distribution of the time between the moment of infection of a randomly chosen infective and that of his/her infector.

In what follows we will write $R_0f_G(s)$ instead of $\beta (s)$.

Let $i(t)$ denote the expected incidence at $t$ (time since the start of the outbreak), i.e.\ the average community rate of new infections. Since we assume that individuals infected $s$ time units ago (at time $t-s$ if present time equals $t$) will infect new individuals at rate $R_0f_G(s)$ we have the following renewal equation for $i(t)$ (see e.g.\ Diekmann et al., 2013, p212):

\begin{equation}
i(t)=\int_0^t i(t-s) R_0 f_G(s)ds + R_0f_G(t) =\int_0^t R_0f_G(t-s) i(s)ds + R_0f_G(t). \label{i-def}
\end{equation}

The additive term in the above equation derives from the initial infective that is supposed to have started the outbreak at $t=0$. 
It is well known that the incidence $i(t)$ will quickly  approach exponential growth $i(t) \sim Ce^{rt}$
where $r$ is the so-called Malthusian parameter defined as the solution to the Euler-Lotka equation

\begin{equation}
1=\int_0^\infty e^{-rt} R_0 f_G(t)dt. \label{Malthus}
\end{equation}

\begin{figure*}[h]
\begin{tabular}{cc}
\hspace{-1cm}
\includegraphics[scale=0.4]{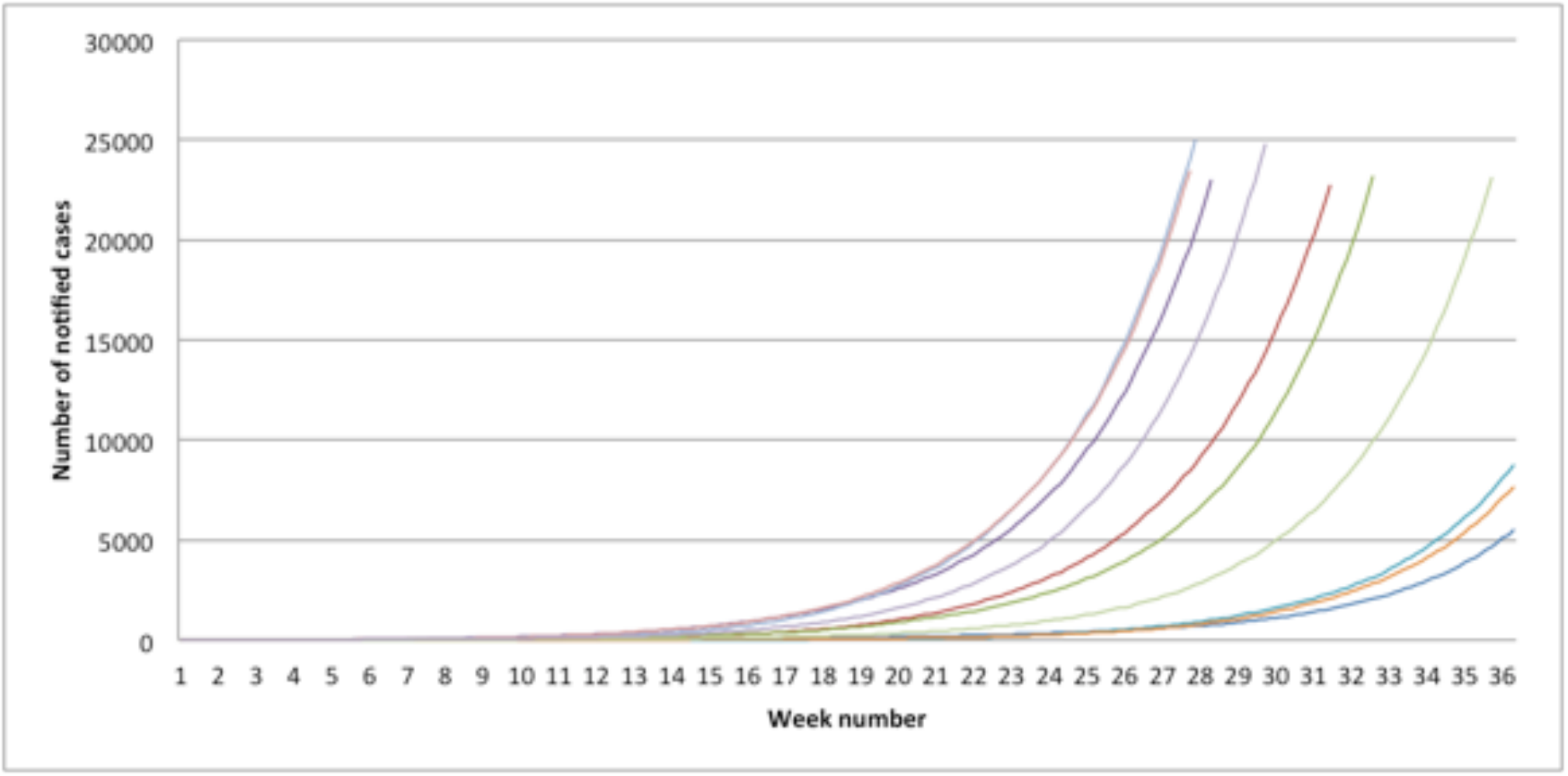}
&
\hspace{-1cm}
\includegraphics[scale=0.4]{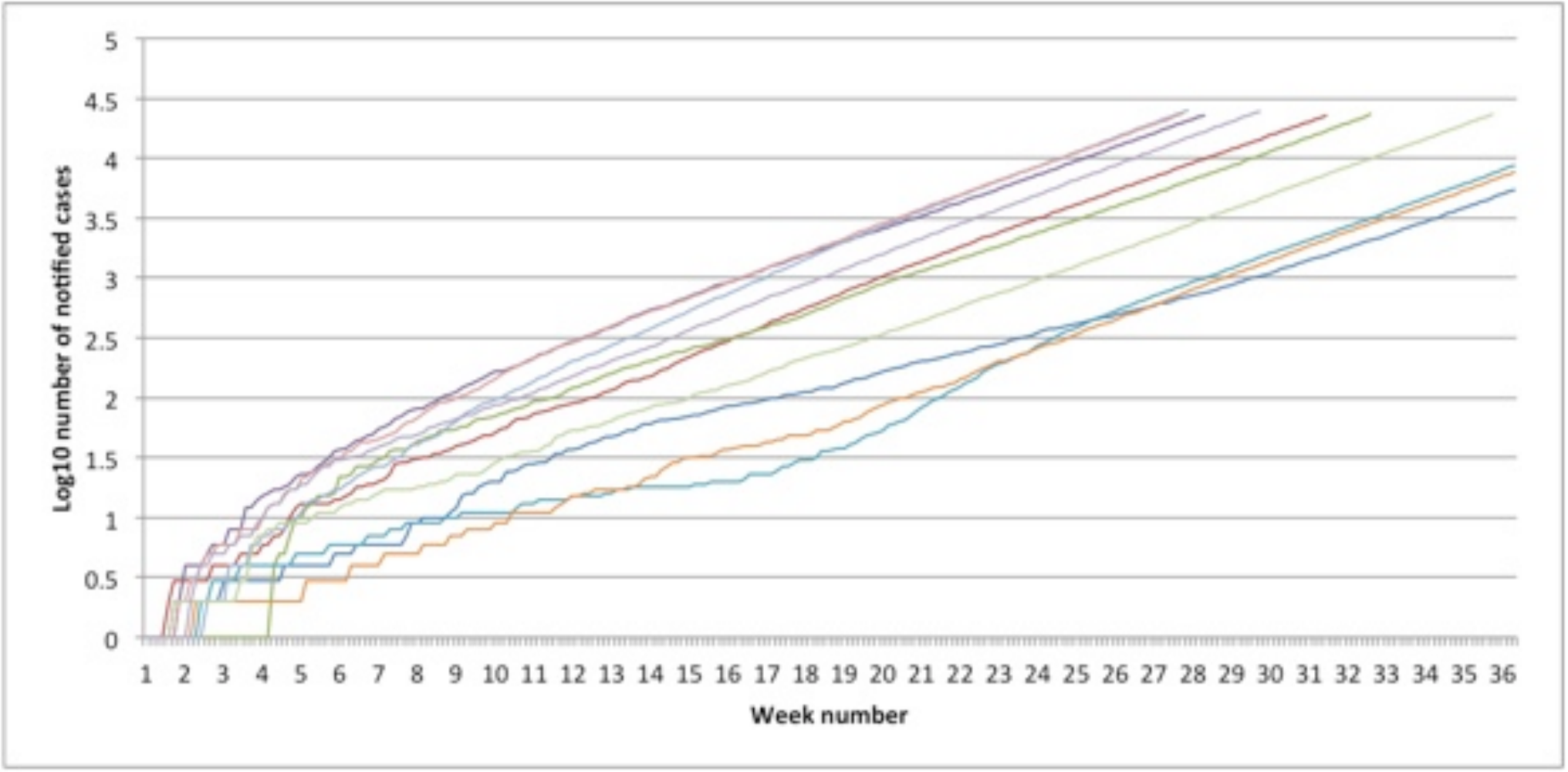}
\end{tabular}
\caption{Initial stages of ten simulated epidemic outbreaks with $R_0=1.7$ and generation time distribution $G$ being Gamma distributed with mean = 15 days and standard deviation = 8.7 days. Incidence over time in original and log-scale.}\label{fig_simul}
\end{figure*}

To simplify matters, we will assume that this exponential growth of new cases holds from the start. The validity of this assumption will be shown by subsequent simulations. In Figure \ref{fig_simul}, ten simulated epidemics are plotted over time showing the exponentially growing feature (clearly visible on the logscale).

Thus,  knowing the generation time distribution $f_G(\cdot)$ and one of $R_0$ and $r$ allows the determination of the other one (cf.\ Wallinga and Lipsitch (2007)). For this reason, estimation of the generation time distribution $f_G(\cdot)$ becomes paramount in this model formulation and will be extensively discussed in following sections. Also, various rather general conclusions about the effects of varying the components of (\ref{Malthus}) related to the directions of biases in the estimation of these components can be drawn. The mathematical details are given in the Supplementary Information and the specific results will be discussed in the relevant sections.

In the model description above, the expected incidence $i(t)$ was a time-continuous deterministic function. The true incidence is, of course, integer-valued and, in most situations, observations are not made continuously but are aggregated in discrete time units such as days or weeks. A related discrete time model is obtained by  suitably discretizing Equation (\ref{i-def}) so that the expected incidence $I(t+1)$ in time (interval) $t+1$ is expressed as 
\begin{equation}
I(t+1)=\sum_{s=1}^t I(t+1-s)R_0p_G(s)=\sum_{s=1}^t R_0 p_G(t+1-s)I(s) ,\label{disc-renewal}
\end{equation}
where $p_G$ is a discrete probability distribution for the generation time corresponding to the continuous time distribution $f_G$. A natural statistical model for data collected daily or weekly is then to assume that the number of newly infected $I(t+1)$, given previous incidence, follows a Poisson distribution with mean parameter $\sum_s R_0 p_G(t+1-s)I(s)$ (cf.\ Team WER et al., 2014).

Finally, the quantitative evaluation of many theoretical results requires explicit assumptions about the involved probability distributions and other parameters typical of the disease under study. As illustrations, we have chosen to use Gamma distributions, where possible, because of their flexibility and analytical properties, and parameters compatible with the recent 2014 West Africa Ebola outbreak (cf.\ Team WER et al., 2014). Various formulae related to these assumptions are collected in Supplementary Information.

\section{Looking backwards rather than forwards in time}\label{sec-backwards}

The generation time distribution $f_G(t)=\beta (t)/R_0$ describes the variability of the (random) time between the moment of infection of an individual and the moments that this individual infects other individuals (so an individual who infects three people gives rise to three generation times). When trying to estimate this quantity from outbreak data, the most common situation is where infected cases are contact-traced, i.e.\  the infectors of cases are identified, and the duration between the infection times of  infector and infectee is ascertained (in theory, but see also next section). This seemingly innocent choice of looking backwards rather than forwards in time (measuring duration backwards from an infectee rather than forwards from an infector) actually modifies the distribution of observed times in the early stage of an outbreak when the epidemic grows at an exponential rate (see e.g.\ Svensson (2007), Scalia Tomba et al (2010), Champredon \& Dushoff (2016)). 
The reason is that, by looking backwards in time, long generation times will be underrepresented and short generation times will be overrepresented because exponential growth implies that there are many more recently infected individuals who are potential infectors compared to those infected longer ago. As a consequence, if the generation time distribution is estimated from a sample of backward generation times, the resulting distribution  $ f_B(s)$ will be different from the true generation time distribution $f_G(s)$.

In fact, it can be shown that (see the above references) the backward generation time has density $f_B(t)=e^{-r t}R_0f_G(t)$ (note that Equation (\ref{Malthus}) implies that this function integrates to 1). It can also be shown that this density has smaller mean than $f_G(\cdot)$ (see Supplementary Information). We can in fact say more. If the backward generation time distribution is used to calculate the exponential growth rate in Equation (\ref{Malthus}), assuming that the correct value of $R_0$ is used, the resulting growth rate $r_B$ will always be larger than $r$. The exact relation is model specific, but as an example one may consider the simple Markovian SIR model, where the infectious period has an exponential distribution with expected value $1/\gamma$ and the infectious contacts, in the initial phase of the epidemic, occur with intensity $\beta$ during the infectious period. The resulting $R_0$ is $\beta/\gamma$, $r=\beta - \gamma$, $f_G(t)=\gamma e^{-\gamma t}$ and $f_B(t)=\beta e^{-\beta t}$. Then, the resulting $r_B$ equals $r_B=R_0 r$. With typical values of $R_0$ being between 1.5 to 2, this means that the exponential growth rate will be grossly overestimated (50-100\%), when using Equation (\ref{Malthus}).

One can also predict the effect on estimating $R_0$ of using $f_B(t)$ instead of $f_G(t)$ assuming that the growth rate $r$ is known or approximately known through observations (see Supplementary Information). Since incidence essentially is $R_0$  $\times$ a weighted sum of previous incidence (c.f.\ Equation (\ref{disc-renewal})) and $f_B(t)$ attributes too much weight to recent incidence (shorter generation times), which is higher than earlier incidence, there will be a compensatory underestimation of $R_0$. In Section \ref{sec-illustration}, where we compute biasing effects with parameters inspired by the recent Ebola outbreak, we illustrate both scenarios: estimation of $R_0$ when $r$ is estimated directly from data, and estimation of $r$ if instead $R_0$ is assumed known, for example from earlier outbreaks or case studies.

\section{Replacing Generation times with Serial intervals}\label{sec-serial}

As described earlier, the generation time is defined as the time between moments of infection of an infector-infectee pair. However, in real life, the infection times are rarely known. Instead, typically, the onset of symptoms is observed. For this reason, the serial interval, which is defined as the time between symptom appearance in the two individuals mentioned above, is often used as a surrogate for the generation time.

We now study the effects of using serial intervals instead of "true" generation times when estimating the generation time distribution $f_G(\cdot)$ and on derived quantities, such as $r$ and $R_0$.

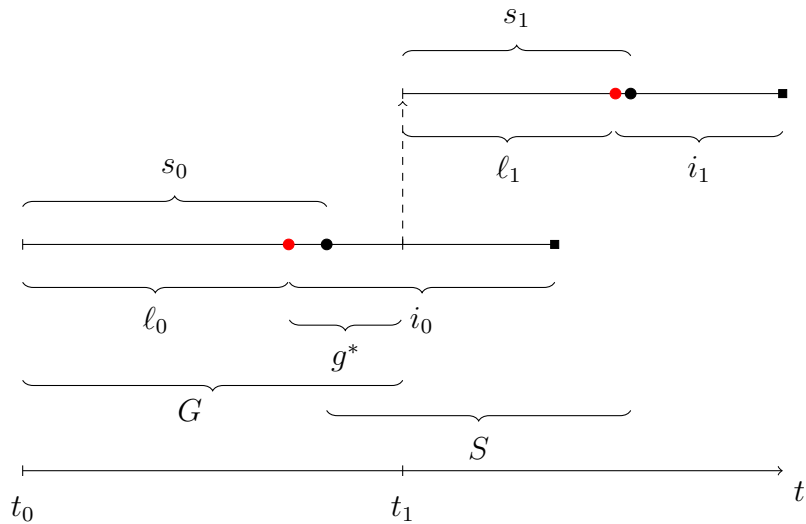
\begin{figure}[h]
\centering
\begin{tikzpicture}
% draw the infectee arrow
 \draw[-](5,5)--(10,5);
  % draw the tick marks at infection
 \draw[shift={(5,5)},color=black] (0pt,2pt) -- (0pt,-2pt) ;
  \draw[shift={(5,3)},color=black] (0pt,2pt) -- (0pt,-2pt) ;
  %circle and arrow of infectee
  \filldraw ([xshift=-1.5pt,yshift=-1.5pt]10,5) rectangle ++(3pt,3pt);
\filldraw (8,5) circle (2pt);
\filldraw[red] (7.8,5) circle (2pt);
% draw the infector arrow
 \draw[-](0,3)--(7,3);
  % draw the tick mark at infection of infector
  \draw[shift={(0,3)},color=black] (0pt,2pt) -- (0pt,-2pt);
  %circle and arrow of infector
  \filldraw ([xshift=-1.5pt,yshift=-1.5pt]7,3) rectangle ++(3pt,3pt);
\filldraw (4,3) circle (2pt);
 \filldraw[red] (3.5,3) circle (2pt);
 \draw[->](0,0)--(10,0) node[anchor = north west] {$t$};
   % draw the tick marks on time axis
  \draw[shift={(0,0)},color=black] (0pt,2pt) -- (0pt,-2pt);
  \draw[shift={(5,0)},color=black] (0pt,2pt) -- (0pt,-2pt);
  \node at (0,-0.5){$t_0$};
  \node at (5,-0.5){$t_1$};

% draw dashed horizontal for infection 
\draw[-, dashed, ->](5,3)--(5,4.9);
 
\draw[decorate,decoration={brace,amplitude=4pt,mirror}] 
 (0,2.5)  -- (3.49,2.5); 
\node at (1.75,2){$\ell_0$};

\draw[decorate,decoration={brace,amplitude=4pt,mirror}] 
 (3.51,2.5)  -- (7,2.5); 
\node at (5.25,2){$i_0$};

\draw[decorate,decoration={brace,amplitude=4pt,mirror}] 
 (3.51,2)  -- (4.98,2); 
\node at (4.25,1.5){$g^*$};

\draw[decorate,decoration={brace,amplitude=4pt,mirror}] 
 (0,1.2)  -- (5,1.2); 
\node at (2.2,0.8){$G$};

\draw[decorate,decoration={brace,amplitude=4pt,mirror}] 
 (4,0.8)  -- (8,0.8); 
\node at (6,0.3){$S$};

\draw[decorate,decoration={brace,amplitude=4pt}] 
 (0,3.5)  -- (4,3.5); 
\node at (2,4){$s_0$};

\draw[decorate,decoration={brace,amplitude=4pt}] 
 (5,5.5)  -- (8,5.5); 
\node at (6.5,6){$s_1$};

\draw[decorate,decoration={brace,amplitude=4pt,mirror}] 
 (5,4.5)  -- (7.75,4.5); 
\node at (6.4,4){$\ell_1$};

\draw[decorate,decoration={brace,amplitude=4pt,mirror}] 
 (7.8,4.5)  -- (10,4.5); 
\node at (8.9,4){$i_1$};

% \end{center}
\end{tikzpicture}
\caption{Relation between generation time $G$ and serial interval $S$. The red circles indicate end of latent period and start of infectious period, the black circles indicates onset of symptoms, and black boxes end of infectious period (either by death or recovery). In the figure, the infectious period starts slightly before onset of symptoms, but, in general, the relation between these times is disease dependent.} \label{Gen_Ser_fig}
\end{figure}

Considering the disease and infectivity history of an individual, starting from the moment of infection, several time periods are of interest (see also Figure \ref{Gen_Ser_fig}). We denote the time of infection of this individual by $t_0$, there may be a latent period of length $\ell_0$ until start of infectivity followed by an infectious period of length $i_0$, and a time from infection to symptoms (incubation period) of length $s_0$. Assume that another individual is infected by the first one after a time $g_*$ within the infectious period $i_0$, i.e.\ at time $t_1=t_0+\ell_0+g_*$ and that this second individual shows symptoms at time $s_1$ after infection. Then, the generation time is $G = (t_0+\ell_0+g_*)-t_0$ and the serial interval $S=(t_1+s_1)-(t_0+s_0)$, see Figure \ref{Gen_Ser_fig} for an illustration.

Although much work has been devoted to estimating the distributions of incubation, latency and infectious periods for various diseases, relatively little has been done regarding their joint distribution. Let us only assume, for a start, that the involved times are independent between different individuals and that corresponding periods have identical marginal distributions between different individuals. We may then rewrite the above expressions as 
\begin{equation}
G =s_0+( \ell_0+g_*-s_0)\quad\text{ and }\quad S=s_1+(\ell_0+g_*-s_0).\label{Gen_Ser}
\end{equation}

These representations are quite unnatural, but show the common structure of $G$ and $S$. For instance, we see that $S=G+(s_1-s_0)$ and thus the expected values of $G$ and $S$ will be equal since $s_0$ and $s_1$ are assumed to have identical expected values. We also see in Equation (\ref{Gen_Ser}) that $S$ is the sum of two independent components (since they regard different individuals) and thus its variance will be the sum of the variances of these components, while the variance of $G$, in addition to the same sum of two variances,  will also contain the term $+2\mathrm{Cov}(s_0, \ell_0+g_*-s_0)$, by the rule for variances of sums. Depending on assumptions, we can now have different results. If we assume that $g_*$, the time from the start of the infectious period to infection, depends only on the duration $i_0$ of the infectious period (e.g.\ $g_*$ could be uniformly distributed in $[0,i_0]$) and that the incubation period $s_0$ is independent of latency period $\ell_0$  and the infectious period $i_0$, then the above covariance term is clearly negative and the variance of $S$ would be larger than the variance of $G$. We can also see that, as assumed in (Team WER et al., 2014), if $\ell_0=s_0$ (the end of latency period/start of infectious period is identical with onset of symptoms), then $G=s_0+g_*$ and $S=s_1+g_*$. It is then true that $G$ and $S$ will have the same distribution, but only if $s_0$ is independent of $g_*$, i.e.\ of $i_0$. Even if this assumption is correct, if the equality is not exact, but $s_0=\ell_0+\delta$, the covariance term above becomes $\mathrm{Cov}(\ell_0+\delta,g_*-\delta)$, which is negative and then, again, $V(S)>V(G)$.

Although it is theoretically possible to have different result, all existing models that we have checked lead to $S$ having equal or larger variance than $G$. Thus the observed serial interval distribution will typically not be an unbiased estimate of the generation time distribution and will have a larger variance. The effects of using a distribution with, in theory, equal mean but larger variance than the true one are model dependent, but typically lead to underestimation of $R_0$, given r, and overestimation of r, given $R_0$ (see Section \ref{subsec-serial} for numerical illustrations and Supplementary Information for analytical results in a specific parametric model).

\section{Multiple exposures}\label{sec-multiple}

Contact tracing means that reported cases are investigated to find out when they have been in contact with infectious individuals, with the aim of finding who the infector was and when the case was infected. If the same procedure has been applied to the infector, a generation time can be calculated. Since the onset of symptoms is known for both cases, usually being the reason for notification, the serial interval can be estimated and, also,  the respective incubation periods (the time between infection and onset of symptoms) determined. In practice, when infected individuals are contact traced, certain cases will have one unique possible infection time, but others will have several potential infectors or infection occasions, or no identified exposure. In the first situation, it is clear who  the infector was and also how long  the incubation period was, and, in the last case, when there is no identified exposure, there is not much to do. But, in the second scenario, it could be any one of the potential exposures that caused the infection, also implying that the incubation period could be one out of several values. In the current section we describe how to infer the incubation period distribution of contact traced individuals in this situation, and also to study  the effects of not correctly acknowledging the multiple exposures situation. It should be noticed that this is not a standard problem in survival data analysis, where it is usually assumed that the time origin of durations is well defined. There is apparently quite little in the literature about the "uncertain origin problem", most of which was done during the 1980's in connection with inference on AIDS data, where the moment of infection of patients was usually not known exactly (see e.g.\ Struthers \& Farewell (1989) or De Gruttola \& Lagakos (1989)). Since the uncertainty about the moment of infection does not only depend on the distribution of the duration under study, as is the case with the more commonly studied censoring situations, but on other factors, a second distribution, for the moment of infection,  has to be introduced (and maybe estimated). In the AIDS case, the uncertainty about the moment of infection was usually restricted to a given time interval (e.g.\ between two blood tests) and the distribution within this interval was, in various approaches, assumed to be uniform, exponentially weighted or general in calendar time. However, this kind of approach is not appropriate to the kind of data that we expect with contact tracing during the early spread of an infection.

Let us start by considering the problem of estimating the incubation period distribution, i.e.\ the time from infection to symptoms, and formulate an appropriate likelihood. Consider  one infected individual with onset of symptoms at time $s$ that has been traced for previous infectious contacts and assume that these exposures took place at the time points $e_1,\dots ,e_k$ where $e_1\le \dots \le e_k < s$. In order to obtain a likelihood we introduce some notation and assumptions. Suppose that at time $t$, the rate of infection exposure equals $\lambda(t)$,  and that the probability of infection upon exposure equals $p$. Finally,  let $g(t)$ denote the density distribution of the incubation period. The likelihood for the infected individual with exposures at times $e_1,\dots ,e_k$  and onset of symptoms at $s$ is given by
\begin{equation}
L(e_1,\dots ,e_k, s) = e^{-\int_0^{s}\lambda(u)du}\prod_{i=1}^k \lambda (e_i)\times \sum_{i=1}^k p(1-p)^{i-1}g(s-e_i). \label{single-lik}
\end{equation}
 We will discuss the estimation problem arising from Equation (\ref{single-lik}) below, but we start with some general considerations.
 
One can imagine several ways to try to avoid the multiple exposures problem, all however leading to biased estimation. One approach could be to pretend that the earliest potential infector is the infector, so that the likelihood contribution related to the incubation time distribution is changed to $g(s-e_1)$ (this would approximately be the same as Equation (\ref{single-lik}) if $p\approx 1$). This would however certainly lead to the duration of incubation periods being \emph{overestimated}. The opposite approach, to pretend that the most recent contact was the infector, would similarly lead to  \emph{underestimation}. A type of compromise could be to treat all potential contacts as being potential infection times (to different cases). As a consequence, one observation with $k$ multiple potential infectors would then result in $k$ \emph{independent} incubation periods $s-e_1,\dots, s-e_k$, and the likelihood contribution would become $\prod_{i=1}^kg(s-e_i)$. Compared to a correct analysis using Equation (\ref{single-lik}), where the shorter incubation periods are given relatively lower weight due to the factor $(1-p)^{i-1}$, such an analysis would lead to the incubation periods (and serial intervals) being \emph{underestimated} (and precision of the biased estimate being overestimated because of the apparent higher number of data points). A related assumption, leading to the same conclusion, would be to assume that the infection time is uniformly distributed among all potential exposures (which would approximately hold true if $p\approx 0$).

An alternative approach to overcome the difficulty of having multiple potential infectors, is to base inference only on individuals having one exposure, i.e.\  simply leaving out all contact traced individuals having more than one exposure. This clearly increases uncertainty by using less data points. However, it also leads to biased estimates, as we now explain. Individuals having only one exposure and then symptoms must have been infected at this first exposure and thus their infection history is certain. However, the fact that no other exposures have happened during the incubation period favours shorter than usual intervals. In fact, the observed time interval will be the minimum of a typical "inter-exposure time" and an incubation time, and will thus have a distribution different from a generic incubation time.  In order to obtain explicit expressions for the size of the bias, explicit models of the "exposure process" and the incubation time distribution are required.

However, the above discussed approaches are not necessary. The correct way of analysing this type of data is to use the likelihood (\ref{single-lik}). It is reasonable to condition on the number and times of exposure, since these essentially depend on the "inter-exposure process", and to base inference on the second part of the likelihood expression only, containing parameters $p$ and the incubation distribution $g$. Assuming a parametric form for $g$, e.g.\ a two-parameter gamma distribution, the problem is non-standard but essentially a three-parameter maximum-likelihood problem with natural bounds on parameters.

It is also possible to find nonparametric (distribution-free) moment estimators of $p$, the mean and the variance of the incubation time at the cost of assumptions about the contact process, e.g.\ as a constant rate Poisson process. Details about one set of such moment estimators and their performance are given in Supplementary Information.

\section{Case fatality rate}\label{sec-fatality}

Estimation of the case fatality rate $p$, i.e.\ the probability of dying as a consequence of an infection, is complicated by the finite time horizon of early data, but also by the exponential increase of cases. As a consequence, the natural estimator of $p$, $D_{obs}/K$, where $K$ denotes the observed number of cases and $D_{obs}$ the observed number of case fatalities up to a given point in time $T$, is biased downwards, since not all deaths that will occur among the $K$ cases have yet occurred at the given time, as noted by e.g.\ Garske et al.\ (2009), Nishiura et al.\ (2009) and Kucharski \& Edmunds (2014). 

However, some general results on delayed observations (see Supplementary Information) can be applied to the situation and an estimate of the bias can be obtained and thus used to correct the estimate. Assume that the fraction $p$ of the $K$ observed cases in the time interval $[0,T]$ is expected to die, but the death of each one of the deadly cases is "delayed" relative to the moment of notification by a time, the time from notification to death, that we assume has a probability density $h(s)$ with cdf $H(s)$. Since the number of notified cases is expected to grow exponentially at rate r,  the fraction $D_{obs}/K$ is expected to be close to $p\pi (T)$, where
$$ 
\pi(T) = \int_0^T re^{-r(T-s)}H(T-s) ds,
$$
which, in turn, will be close to
$$ 
\pi(\infty) = \int_0^\infty e^{-rs}h(s) ds,
$$
where $r$ is the exponential increase rate of notified cases. Thus knowledge of $r$ and of the distribution $h$ can be used to correct the naive estimate $D_{obs}/K$. As an illustration, if the distribution $h$ is assumed to be a simple exponential distribution with expected value $\mu$, then $\pi(\infty) = \frac{1}{1+r\mu}$.

In (Team WER et al., 2014) another approach is used, namely estimating only on cases with a known final destiny (death or recovery). Let us denote by $R_{obs}$ the number of cases who are observed as recovering in [0,T] and, as before, by $D_{obs}$ those that have died. Then, another application of the general delayed events formulae gives that the fraction $R_{obs}/K$ will be close to $(1-p)\rho(T)$, where, letting $M(t)$ denote the cdf of the time from notification to recovery, with corresponding density $m(t)$,
$$ 
\rho(T) = \int_0^T re^{-r(T-s)}M(T-s) ds,
$$
which, in turn, will be close to
$$ 
\rho(\infty) = \int_0^\infty e^{-rs}m(s) ds.
$$
The estimator $D_{obs}/(D_{obs}+R_{obs})$ will then be close to 
$$
\frac{p\pi(\infty)}{p\pi(\infty)+(1-p)\rho(\infty)}=\frac{p}{p+(1-p)\frac{\rho(\infty)}{\pi(\infty)}}.
$$
Thus, the estimate will be (approximately) unbiased only if $\rho(\infty)=\pi(\infty)$. If $\rho(\infty) < \pi(\infty)$ then the CFR will be overestimated. This happens if the time to remission is stochastically larger than the time to death, which is the case for many diseases, for instance this seems to be the case for Ebola (see Section \ref{subsec-cfr}). However, the reverse case, i.e.\  $\rho(\infty) > \pi(\infty)$, is also interesting, e.g.\ for influenza (see Garske et al.\ (2009)).

 \section{Results}\label{sec-illustration}

We now illustrate our findings based on data and estimates from the recent Ebola outbreak in West Africa as described in Team WER et al.\ (2014). We emphasize that the results are not based on raw data and only use convenient approximations of the estimates obtained in the paper as  plug-in estimates to illustrate the magnitude of the various potential biasing effects in a realistic parameter setting. We also report results from simulations using the same parameter setting. Details about  theoretical derivations and about the simulation program and related results are reported in Supplementary Information. The Sections \ref{subsec-back} and \ref{subsec-serial} respectively illustrate the biases arising from the use of backward instead of forward generation times, and the use of serial intervals instead of generation times in Equation (\ref{Malthus}) to estimate $R_0$ and $r$. In  Section  \ref{subsec-multiple}, the effects of using data from individuals with only one possible infector, instead of complete data with multiple possible infectors, are studied. In Section \ref{subsec-cfr}, some results related to the estimation of the case fatality rate are derived, while Section \ref{subsec-sim} contains some interesting observations obtained from the simulated epidemic outbreaks. 

\subsection{Looking backwards}\label{subsec-back}

We assume that the generation time follows a gamma distribution $G\sim \Gamma (\alpha, \lambda)$ with $(\alpha,\lambda) = (3,0.2)$ and that $R_0 = 1.7$. For given basic reproduction number $R_0$ this induces a true exponential growth rate equal to $r=\lambda(R_0^{1/\alpha}-1)$. The generation time when looking backwards in time (by means of contact tracing reported cases) also follows a gamma distribution, but with different parameters: $B\sim \Gamma (\alpha, \lambda + r=\lambda R_0^{1/\alpha}$). If the exponential growth rate is computed for this (backward) generation time distribution and the true $R_0$, we get $r_B= R_0^{1/\alpha}r$. Conversely, if we assume that the true $r$-value is known and we compute the corresponding basic reproduction number, we get $R_0^{(B)} = (1-(\frac{r}{\lambda+r})^2)^\alpha R_0$.

In numbers, our assumptions about the generation time distribution and $R_0$ correspond to an expected value of 15 days and standard deviation (sd) = 8.66 and exponential growth rate $r=0.0387$ (per day). The backward generation time will instead have mean 12.6 days and $sd = 7.26$. The induced exponential growth rate then equals $r_B=0.0462$. Thus, the growth rate estimate is  overestimated by 19\%.  Conversely, if the true value  $r=0.0387$ is used in Equation (\ref{Malthus}), but $\alpha$ and $\lambda$ are taken from the backward generation time distribution, the result is $R_0^{(B)} = 1.57$ as compared to the true value $R_0=1.7$, and $R_0$ will be underestimated by 8\%.

\subsection{Serial intervals}\label{subsec-serial}

We start by looking at the consequences of overestimating the variance of the generation time distribution by using serial intervals instead of generation time data in the simplified framework where both distributions are of the Gamma type and the difference is represented by the coefficient of variation of the serial interval distribution being larger than that of the generation time distribution by a factor $c>1$, while the means are equal, as predicted by theory (see Section \ref{sec-serial}).
If we assume the same basic parameter values as in the preceding Section (i.e.\ the generation time follows a gamma distribution $G\sim \Gamma (\alpha, \lambda)$ with $(\alpha,\lambda) = (3,0.2)$ and $R_0 = 1.7$), and calculate the biases resulting from e.g.\ $c =$ 1.1, 1.2, 1.5 and 2, we find that the corresponding $R_0$ values, assuming the true $r$ value, are downbiased by 0.9, 1.8, 4.8 and 9.6\%, respectively, while the corresponding $r$ values, assuming $R_0 = 1.7$, are overestimated by 1.9, 4.1, 12.3 and 32.9\%, respectively. Thus, sizeable bias can be obtained if the serial intervals are much more variable than the generation times.

In Team WER et al.\ (2014), the generation time distribution was estimated from observed serial intervals, under the assumption that the distributions would be equal, which was assumed to be a consequence of the exact coincidence of appearance of symptoms and beginning of infectious period (i.e.\ exact equality of latent period and incubation time, see Section \ref{sec-serial}). In our simulations, we however assume that the equality is only approximately exact by taking the incubation period equal to a factor U times the latent period, where U is chosen uniformly  in the interval [0.8,1.2], thus assuming that the difference is at most $\pm 20\%$, with mean $0\%$.. Thus symptoms are allowed to appear a little before or after the start of the infectious period. Straightforward calculations yield that this modification corresponds to a c value of 1.026, i.e\ a very small increment in variability of serial intervals relative to generation times, which would modify $R_0$ and $r$ values calculated as above by only 0.2 and 0.5\%, respectively.

Thus, the above favourable assumptions combined with the parameter values as estimated by Team WER et al.\ (2014) lead to a very small effect of the use of serial intervals instead of generation times, but we would like to point out that the chosen model for the relation between these quantities is a very specific one, chosen to avoid statistical complications. In other situations, larger differences between latent periods and incubation times, or other assumptions about the time order of events in the natural history of the infection, may lead to larger differences between serial intervals and generation times.

\subsection{Multiple exposures}\label{subsec-multiple}

In order to estimate the effects of basing the estimates of durations on individuals having only one exposure (see Section \ref{sec-multiple}), some additional assumptions are needed. For the recent Ebola epidemic, Team WER et al.\ (2014) find that the incubation period distribution $f_D$, assumed to be equal to that of the latent period, is Gamma distributed with mean $11.4$ days and $sd=8.1$ days and that the serial intervals are Gamma distributed with mean 15.3 days and $sd=9.3$ which is also assumed to be the distribution of generation times. They also report that approximately 25\% of the contact traced individuals had one unique infector and 75\% had more than one potential infector. We will use the above parameter values for this example. With the complete data, it would have been possible to estimate the contact rate $\lambda$ and the probability $p$ to get infected by a close Ebola contact separately. Here we can only use that 25\% had a single contact. We simply assume that $p=0.5$ and equate $P(\text{a single contact})=p\int_0^\infty e^{-\lambda s}f_D(s)ds$ to the empirical value 0.25. The result is that $\lambda=0.0725$ per day (so about one close contact every 2 weeks for the contact-traced individuals).

Once values for $p$ and $\lambda$ are available, one can compute the mean incubation period for observations having only one possible infector: 
$$
E(D|\text{one possible infector})=\int_0^\infty spf_D(s)e^{-\lambda s}ds/P(\text{one possible infector})\approx 8.1.
$$
Thus, the mean incubation period for infectees with only one potential infector will be $11.4-8.1=3.3$ days shorter than the mean incubation period had all observations been used. This in turn implies that the mean generation time from the same data would  be underestimated by 3.3 days, giving a mean of 12 instead of 15.3 days. Assuming that the standard deviation remains unchanged ($= 9.3$ days), the estimated generation time distribution would be Gamma distributed with mean 12 and $sd=9.3$. Assuming $R_0=1.7$, the "true" exponential growth rate equals $r=0.0383$, whereas the exponential growth rate estimated from the contact traced individuals having only a single unique infector  would approximately equal $r_{single}=0.0522$, which overestimates the true value by 36\%. Once again, assuming $r=0.0383$ to be known (e.g.\ estimated from the observed growth rate), instead leads to $R_{0single}=1.50$, an underestimation of 12\%.

Instead, going back to the use of maximum likelihood estimation based on (\ref{single-lik}) and  an alternative set of moment estimators, we have simulated observations from 500 individuals (see Supplementary Information for details), showing that estimates of $p$ and the parameters of the incubation period distribution seem reasonably unbiased, given the parameter setting and assuming the correct distributional form in the likelihood method. If the incubation period has a  distribution differing from the assumed (gamma) model distribution (the log-normal distribution, in our simulation), the moment-estimators still perform well, but the maximum likelihood estimates of mean and variance derived under the assumption of gamma distributed incubation times now acquire some bias. The speed of convergence of estimates and further properties under misspecification of assumptions need further study, but this initial experiment shows that unbiased estimation based on all observations is possible.

\subsection{Combined effect of generation time biases}\label{subsec-combined}

The bias effects of the three sources of errors as well as the combined effect, is summarized in Table \ref{Tab_error} below. The combined effect is obtained assuming that that the three sources of error act independently.
\begin{table}[ht]
\centering
\caption{Bias in estimating $R_0$ assuming $r$ known, and vice versa, using Equation (\ref{Malthus}) for three sources of errors discussed in the text. Parameter values and other assumptions are taken from the Ebola outbreak (Team WER et al., 2014). See text for further explanations. }
\vskip.3cm 
\begin{tabular}{|l| c c |}
\hline
 Source of error & Bias in $R_0$ & Bias in $r$ \\
 & given $r$ & given $R_0$\\
 \hline
Looking backwards & -8\% & +19\% \\
Serial times & -0\% & +0\% \\
Multiple exposures & -12\% & +36\% \\
Combined effect & -20\% & +62\% \\
\hline
 \end{tabular}\label{Tab_error}
\end{table}

\subsection{Case fatality rate}\label{subsec-cfr}
 Team WER et al.\ (2014) report that the average time from symptoms to death is $5+4=9$ days, while to remission the average time is $5+12=17$ days. The numerical consequences of the results in Section \ref{sec-fatality} can be sizeable. If we assume that we have exponential growth with a doubling time of say 20 days, the growth rate $r$ becomes 0.0347. Under the simplifying assumption that the time from notification to death follows an exponential distribution with mean $\mu = 9$ days, the multiplier $\frac{1}{1+r\mu}$ becomes 0.76, i.e there is an underestimation of about 24\% of the CFR, using the simple estimator. However, similarly the factor $\rho(\infty)$ will be 0.63 and, assuming a CFR of 70\%, say, the estimator relying only on cases with a known final destiny will overestimate the CFR by approximately 5\%. 

\subsection{A simulation study}\label{subsec-sim}
In order to better study the behaviour of various observables during the early phase of  an outbreak, we have conducted simulations and evaluated various statistics (for details of the simulation model and parameters, see Supplementary Information). As before, values were chosen to be similar to the recent Ebola outbreak in West Africa. Only simulations of outbreaks becoming large (reaching at least 4500 cases, this number was chosen being the number of reported cases at which predictions were made in Team WER(2014)) are considered. The most interesting findings are as follows:

\textbf{Time from introduction of first case to 4500 notified}
Average and median approximately 200 days ($sd=33$), but with a range of $[123,358]$.  Since these numbers reflect the time from the introduction of the first case until the day 4500 in total have been notified, they are a couple of days longer than what will be observed from the first day a case is notified. A deterministic estimate would probably be $ln(4500)/r=217$, where $r$ is the exponential increase rate.  The slight difference might be due to the conditioning on non-extinct trajectories.
Furthermore, the largest part of variability is in the first part of the epidemic (cf.\ Figure \ref{fig_simul}). If one divides the time period in a first part until the first 100 (cumulative) cases are reached and a second part until level 4500 is reached, one finds that the first part has average length 102 days with  $sd= 28$ and the 95\% central range $[63,174]$ and that the second part has mean 98 days, $sd=10$ and 95\% range $[83,119]$.

\textbf{Stability of various ratios when 4500 cases have been notified}
At the time of 4500 cases notified (in total), the numbers of individuals  infected, who had died or who had gotten well, were recorded. It should be noted that the total number of infected is not observable, but of interest to estimate. It should also be noted that all numbers above are examples of delayed observations (see Supplementary Information), from infection to notification and from notification to final destiny. One should therefore expect that the ratios should stabilize around values given by the formulae in that Section. The ratio notified/infected was, on average, 0.70 with 95\% of values between 0.68 and 0.72, essentially identical to the theoretical prediction based on "knowing" the incubation time distribution, which is assumed to also be the time from infection to notification. The narrow range of the observed ratios indicates that this is a viable method to predict the true actual size of the outbreak from the number of notified cases, given that good estimates of the distribution of time to notification and exponential increase rate are available. An analogous result holds for the ratios of infected to dead or recovered individuals. With the specific parameters of the simulations, at 4500 notifications, on average 3350 of them had either recovered or died, and the remaining 1150 remained between notification and final destiny. At the time of 4500 notifications, an average of 1900 additional individuals had been infected but not yet notified.

\textbf{Observed generation times and serial intervals}
In each simulation run, 500 generation times and 500 serial intervals were sampled from the first 4500 notified individuals by systematically  taking every ninth individual until the sample was complete. The times and intervals were ascertained "backwards", i.e.\ the infector of the chosen individual was identified and time distance between the respective infection or symptom times was recorded. The distributions of sample means are reasonably concentrated around the respective central values, 12.5 for serial intervals and 12.6 for generation times. It should be remembered that theory predicts that both should have the same expected value which should be less than the true generation time expectation, which is 15. Theory again predicts that the backward generation time should have mean 12.57, which is not far from what is observed. The variance of the true generation time is 75 ($sd = 8.7$) and both variance estimates from the simulation samples tend to be much less, somewhat above 50 ($sd = 7.1$). This also leads to the useful conclusion that serial times are affected by the same "contraction" as generation times when ascertained "backwards", at least in the chosen parameter setting.

\textbf{Predicting the size of the outbreak at a later time}
Finally, we study the performance of the renewal equation (Eq. \ref{disc-renewal} ) approach proposed in Team WER (2014). This approach is intended to allow estimation of $R_t$ (in our simulation, $R_t$ is kept constant  $= R_0$ all the time and the method is adapted accordingly) and to further allow prediction of cases 6 weeks (42 days) after the last observed datum. The results, using probabilities derived from observing backward serial times, thus biased with respect to the true generation time distribution, indicate that this method seriously underestimates $R_0$ but has good predictive power anyway. The 95\% range of the ratio predicted/true values at 6 weeks after the level 4500 notified has been reached is $[-10\%,+12\%]$, which is slightly better than just using a growth rate based estimate (i.e.\ just multiplying with $\exp(42r)$, using some good estimate of the growth rate $r$) derived from the first 4500 cases. This is perhaps natural, since the method amounts to an adaptive regression method which has fitted all the observed data up to level 4500 as well as possible, and then extrapolated this fit.
As predicted by theory, the estimate of $R_0$ that results from this method is downbiased, in fact the true value 1.7 is never reached in 1000 simulations. The average estimate of $R_0$ is 1.57, with 95\% range $[1.51,1.63]$, compared to the true value 1.7, thus having an average bias of $-8\%$.

\section{Discussion}\label{sec-disc}

In the current paper we have, by  means of modelling, analysis and heuristics, both theoretical and simulation-based, in a setting resembling the recent 2014 Ebola outbreak, studied inferential problems in an ongoing epidemic outbreak in its early stage. Our analyses give insights into where biases might "hide" and also how to avoid these biases. We have studied three potential sources of bias: 1) backward estimation of generation times (contact tracing), 2) using serial intervals instead of generation times, and 3) contact tracing leading to several potential infectors thus making infection time uncertain. Importantly, all three sources lead to biases in the same direction, causing the basic reproduction number $R_0$ to be \emph{underestimated} if the epidemic growth rate $r$ is correctly estimated. The converse is also true, namely that the growth rate will be \emph{overestimated} if a correct estimate of $R_0$ is available, but this situation is likely to be less common in practice. 

Using parameter values stemming from the recent Ebola outbreak, it is shown that these biasing effects can be substantial; in magnitude, the third effect (multiple exposures) is largest and the second effect (serial intervals replacing generation times) is smallest. In particular since all three effects lead to bias in the same direction, the combination of their effects can be quite large. If we assume that the biasing effects act independently and take parameters and assumptions from the Team WER (2014) Ebola paper as numerical illustration, then the estimate of $R_0$ could be negatively biased by at least about 10\%, up to 20\%, depending on how estimation is performed. In our illustration, the true $R_0$ was assumed equal to 1.7 and our estimate of potential bias indicates that an estimate could be as low as 1.36. Such a difference can have quite large consequences when planning control measures. For instance, the critical immunization level (both for vaccination and any other measure aimed at reducing infection) is usually calculated as $v_c=1-1/R_0$. For the true $R_0$, this results in $v_c=41\%$, while the lower biased estimate yields $\hat v_c=26\%$. The underestimation of $R_0$ may hence lead to suggested preventive measures that are insufficient to stop the spread.

The focus of the paper has been on studying potential biasing effects originating from a typical set of observables in the initial phase of an outbreak. However, there are also some positive observations. The stability of proportions of individuals in different phases of disease during the increasing phase is one, since quite good estimates of the total number of  infected or not yet notified infected could be made, based on number of dead patients or notified ones, if good information about the related stage duration distributions is available. Another positive observation is that accurate inference in the multiple infector problem seems possible, although more research is needed. Finally, many biases can be understood and corrected for if the sampling situation is correctly modelled. It may be difficult to obtain simple analytical results, but simulation can then reveal the performance of various estimation procedures. 

Of course, there are also many other problems related to data from an emerging outbreak not treated in the current paper, important ones being underreporting and reporting delays, but also batch-reporting of numbers. A rather different type of potential source of bias, also not studied here, is when model assumptions are violated. For example, it has been assumed that there was no individual or society-induced changing behavior during the data collection period, and social or spatial effects on spreading patterns have been ignored. Social structures have been shown to have limited effects for estimation in emerging epidemics (Trapman et al., 2016), but spatial effects (cf.\ Lau et al., 2017) clearly play a role in disease spread, and their effect on parameter estimates is yet to be investigated. Changing behavior probably kicks in early in emerging outbreaks of serious diseases like Ebola, and are hence also important to include in future inferential procedures for emerging epidemic outbreaks.

Still, it is our hope that the results can help improving future analyses of emerging outbreaks and the important efforts to guide health authorities in predictions and identifying possible preventive measures.

\section*{Acknowledgements}

T.B.\ is grateful to the Swedish Research Council (grant 2015-05015) for financial support.

\section*{References}

Anderson, R.M.\ and May, R.M. (1991). Infectious diseases of humans: dynamics and control. Wiley.

Champredon, D. \ and Dushoff, J. (2016). Intrinsic and realized generation intervals in infectious-disease transmission. Proc. R. Soc. B, 282: 20152026.

De Gruttola, V.\ and Lagakos, S.W.\ (1989). Analysis of doubly-censored survival data, with application to AIDS.  Biometrics, 45: 1-11.

Diekmann, O., Heesterbeek, H. and  Britton, T. (2013). Mathematical tools for understanding infectious disease dynamics. Princeton University Press.

Garske, T., Legrand J., Donnelly C.A.\ et al.\ (2009). Assessing the severity of the novel influenza
A/H1N1 pandemic. BMJ, 339:b2840.

Jagers P. (1975) Branching processes with biological applications. John Wiley \& Sons.

Kermack, W.O.\ and McKendrick, A.G. (1927). A contribution to the mathematical theory of epidemics. Proc. Roy: Soc. London. 115: 700-721.  Reprinted as Bull. Math. Bio. (1991). 53:33-55.

Kucharski, A.J.\ and Edmunds, W.J.\ (2014). Case fatality rate for Ebola virus disease in west Africa. The Lancet, 384:

Lau, M.S.Y., Dalziel, B.D., Funk, S., McClelland, A., Tiffany, A., Riley, S., Metcalf, C.J.E., and Grenfell, B.T. (2017) Spatial and temporal dynamics of superspreading events in the 2014?2015 West Africa Ebola epidemic. \emph{PNAS}, \textbf{114}: 2337-2342.

Nishiura, H., Klinkenberg, D., Roberts, M.\ and Heesterbeek, J.A. (2009). Early epidemiological
assessment of the virulence of emerging infectious diseases: a case study of an influenza
pandemic. PLoS One, 4(8): e6852.

Scalia Tomba, G., Svensson, {\AA}., Asikainen, T.\ and Giesecke, J. (2010). Some model based considerations on observing generation times for communicable diseases. Math. Biosci. 223:24-31.

Struthers, C.A.\ and Farewell, V.T. (1989).  A mixture model for time to AIDS data with left truncation and an uncertain origin. Biometrika, 76:814-817.

Svensson, {\AA}.  (2007) A note on generation times in epidemic models. Math. Biosci. 208:300-311.

Svensson, {\AA}. (2015) The influence of assumptions on generation time distributions in epidemic models. Math. Biosci. 270:81-89.

Trapman, P., Ball, F., Dhersin, J.S., Tran, V.C., Wallinga, J., Britton, T., (2016). Inferring $R_0$ in emerging epidemics: the effect of common population structure is small.
Journal of The Royal Society Interface 13 (121), 20160288

Wallinga, J.\ and Lipsitch, M. (2007). How generation intervals shape the relationship between growth rates and reproductive numbers. Proc Roy. Soc. B. 275:599-604.

Team WER et al.\ (2014). Ebola virus disease in West Africa -- the first 9 months of the epidemic and forward projections. N Engl J Med. 371:1481-1495.

\newpage
\setcounter{page}{1}
\setcounter{table}{0}
\setcounter{section}{0}

\section*{Supplementary information}\label{Appendix}
In what follows, frequent reference is made to Gamma distributed quantities. We use the notation $\Gamma(\alpha, \lambda)$ for a Gamma distribution with shape parameter $\alpha$ and scale parameter $\lambda$, having expected value $\alpha/\lambda$ and variance $\alpha/\lambda^2$. Further results related to Gamma distributions are given in Section \ref{app-gamma} in this Appendix.

\section{The simulation model}\label{App-simulation}

\textbf{Model structure}
The simulated epidemic has been constructed to be close to the parameters reported for the recent Ebola outbreak by Team WER et al.\ (2014). Each infected individual follows a stochastic SEIR model with all time periods following Gamma distributions, the time unit being 1 day. The latent period E is assumed to be  $\Gamma(2,1/5)$ ($mean = 10$, $sd = 7.1$) and the infectious period  $\Gamma(1,1/5)$ ($mean = 5$, $sd = 5$). After the infectious period, the individual may either recover, with probability 30\%, or die, with probability 70\%. The time until recovery is assumed to be  $\Gamma(4,1/3)$ ($mean = 12$, $sd = 6$) and the time to death $\Gamma(4/9,1/9)$ ($mean = 4$, $sd = 6$). Furthermore, each individual has an incubation time (time until symptoms) which is assumed to be similar to the latent period, but with some variation. The incubation time is given by E times a uniformly distributed variable in the interval [0.8,1.2]. It is assumed that a case is reported when symptoms arise. Finally, during the infectious period, new cases are produced with rate 0.34/day, resulting in $R_0 =0.34\cdot 5= 1.7$. 

\textbf{Duration of simulations}
Only "exploding" trajectories, corresponding to "big" outbreaks, are kept. Outbreaks start with 1 infected individual and are rejected if they do not reach 4500 reported cases. At the time of reaching this level, some statistics are collected and then the simulation is continued for 6 weeks further. The purpose of this continuation is that a prediction of the final level 6 weeks later is attempted, based on the first 4500 cases. Statistics are based on 1000 accepted trajectories.

\textbf{Programming details}
The simulation program was written in standard C, because of the need to keep links between infectors and infectees, in order to simulate contact tracing, and executed on a desktop computer. Results from the simulations were elaborated using the software R to produce the $R_0$ and $r$ estimates, and the 6 week projections as well as statistical summaries. Random number generation for Gamma distributions with non-integer shape parameter used the algorithm of Phillips \& Beightler (1971).

\textbf{Theoretical results}
It is easily shown (see e.g.\ Svensson (2007)) that the generation time distribution $f_G$ in the above model is $\Gamma(3,1/5)$ ($mean = 15$, $sd = 8.7$) and that the Malthusian parameter is $r =\lambda (R^{1/\alpha}_0-1)=0.0387$. The deterministic doubling time is 17.9 days. The stochastic process as such is a Crump-Mode-Jagers branching process, in which the expected incidence of infections $b(t)$ satisfies (see Jagers (1975)) the renewal equation
$$ b(t) = R_0 \int_0^t b(t-u)f_G(u)du + R_0 f_G(t) .$$
The solution to this equation quickly approaches exponential growth $\sim Ce^{rt}$. The same exponential growth, although with different constants, will also apply to the total number of infected, reported, recovered, dead, etc., individuals.

\section{Delayed observations}\label{app-delay}
Suppose that events occur with expected intensity $\lambda(s)$ (cumulative intensity $\Lambda(s)$) on the time interval $[0,T]$. Assume also that each event is observed after a delay which is distributed according to the density $h(s)$ with cdf $H(s)$. Then the expected number of observed events on [0,T] is
$$
\int_0^T\lambda(s) H(T-s)ds
$$
which constitutes the fraction
$$
\pi(T) = \int_0^T \frac{\lambda(s) H(T-s)}{\Lambda(T)} ds
$$
of the expected number of events on the interval. If the intensity $\lambda(s)$ is constant or even polynomially growing, it can be shown that $\pi(T) \rightarrow 1$ as T grows large, while, if the intensity grows exponentially, i.e.\ $\lambda(s) \sim e^{rs}$, then this fraction quickly approaches (after some simple integration steps)
$$
\pi(\infty) =r \int_0^\infty e^{-rs}H(s) ds = \int_0^\infty e^{-rs}h(s) ds
$$
as $T$ grows large, which is equivalent to calculating the expected value $E(e^{-rD})$, with $D$ having a probability distribution with density $h$. If $D$ has distribution $\Gamma(\alpha, \lambda)$, then $ \pi(\infty) = (\frac{1}{1 + rE(D)/\alpha})^\alpha$. It is interesting to note that this is a decreasing function of $\alpha$, for fixed $r$ and $E(D)$. Thus the case $\alpha=1$, i.e.\ the Exponential distribution, yields the largest possible value among Gamma distributions with given mean.

\section{"Backward" observation of generation times}\label{app-backw}
Observing generation times, i.e.\ the time between infection of one individual and another one infected by the first one, has been discussed by several authors, e.g.\ Svensson (2007), Scalia Tomba et al (2010), Champredon \& Dushoff (2016). In the exponentially increasing phase of a homogeneously mixing model, the distribution of times observed "backwards", starting from a randomly chosen newly infected individual, will have density $f_B(t) = e^{-r_Gt}R_0f_G(t)$, where $f_G$ denotes the generation time distribution, $R_0$ the basic reproduction number of the disease and $r_G$ the related Malthusian parameter (this result is approximate in the sense that the truncation effect of the time origin is disregarded). The parameter $r_G$ satisfies the equation
$$
1=\int_0^\infty e^{-r_Gs}R_0f_G(s)ds
$$
If one solves the Euler-Lotka equation for the Malthusian parameter $r_B$ using the density $f_B(t)$ instead, one obtains
$$
1=\int_0^\infty e^{-r_Bs}R_0f_B(s)ds = \int_0^\infty e^{-r_Bs}R_0^2e^{-r_Gs}f_G(s)ds.
$$
This equation can be rewritten as
$$
1/R_0^2 = E(e^{-r_BT}e^{-r_GT})
$$
where $T$ has the generation time distribution $f_G$. Because both functions of $T$ in the expectation are monotone decreasing, their covariance is positive and thus
$$
1/R_0^2 = E(e^{-r_BT}e^{-r_GT}) \geq E(e^{-r_BT}) E(e^{-r_GT}) = E(e^{-r_BT})\frac{1}{R_0}.
$$
Since $E(e^{-rT})$ is a decreasing function of $r$ and $E(e^{-r_GT})=1/R_0$, and since the above inequality translates to   $E(e^{-r_BT}) \leq 1/R_0$, we have that $r_B \geq r_G$.
More specifically, if the generation time is assumed to have distribution $\Gamma(\alpha,\lambda)$, by direct integration, one finds that $r_B = R_0^{1/\alpha}r_G.$\\
By the same kind of argument, denoting a time with distribution $f_B$ by $T_B$ and one with distribution $f_G$ by $T_G$, one finds that
$$
E(T_B) = \int_0^\infty t R_0 e^{-r_Gt}f_G(t)dt = E(T_G R_0 e^{-r_GT_G}) \leq E(T_G) \times E(R_0 e^{-r_GT_G})=E(T_G),
$$
because the two functions of $T_G$ inside the expectation now have different monotonicity.

\section{The Euler-Lotka equation and $G \sim \Gamma(\alpha,\lambda)$}\label{app-gamma}
If the probability density $f$ in the E-L equation  of the form
$$
1=\int_0^\infty e^{-rs}Rf(s)ds
$$
is a $\Gamma(\alpha,\lambda)$ distribution, the equation becomes 
$$
\frac{\lambda^\alpha}{(\lambda+r)^\alpha} = \frac{1}{R}
$$
and thus $r=\lambda (R^{1/\alpha}-1)$ or $R=(1+r/\lambda)^\alpha$.

It should be noted in the $\Gamma(\alpha,\lambda)$ distribution, the coefficient of variation (ratio of standard deviation to mean) is $1/\sqrt{\alpha}$ and that given the expected value $\mu$ and the variance $\sigma^2$, one has $\alpha=\mu^2/\sigma^2$ and $\lambda=\mu/\sigma^2$.

The above results directly apply to calculating the exponential growth rate if the generation time distribution is $\Gamma(\alpha,\lambda)$ and $R_0$ is known or, viceversa, calculating $R_0$ if the exponential growth rate $r$ is known.

It is also easy to see that if the generation time distribution is $\Gamma(\alpha,\lambda)$ and, denoting the corresponding $R_0$- and $r$-values by $R_0^{(G)}$ and $r_G$, the "backwards" generation time distribution (see Section \ref{sec-backwards}) will be $\Gamma(\alpha,\lambda+r_G)$ and, using this density to solve for the exponential growth rate $r=r_B$, assuming $R_0^{(G)}$ as $R$-value, or solving for $R_0=R_0^{(B)}$, assuming $r_G$ as $r$-value, in the general E-L equation above, yields, after some simplification,
$$
r_B=R_0^{(G)\: 1/\alpha} r_G
$$
and
$$
R_0^{(B)}=\left(1-\left(\frac{r_G}{\lambda+r_G}\right)^2\right)^\alpha R_0^{(G)}.
$$
Finally, if the generation time $G$ has a $\Gamma(\alpha,\lambda)$ distribution and another time $S$ has a Gamma distribution with the same mean but larger variance so that the coefficient of variation of $S$ is larger, by a factor $c>1$, than the coefficient of variation of $G$, $S$ will have a $\Gamma(\alpha/c^2,\lambda/c^2)$ distribution (this situation relates to the problem treated in Sections \ref{sec-serial} and \ref{subsec-serial}). Then, straightforward calculations applied to using this density to solve for the exponential growth rate $r=r_S$, assuming $R_0^{(G)}$ as $R_0$-value, or solving for 
$R_0=R_0^{(S)}$, assuming $r_G$ as $r$-value, in the general E-L equation above, yields, after some simplification,
$$
r_S=\frac{1}{c^2} \frac{R_0^{(G)\: c^2/\alpha}-1}{R_0^{(G)\: 1/\alpha}-1}\:  r_G
$$
and
$$
R_0^{(S)}=\frac
                        {\left(1+\frac{r_G}{\lambda} c^2 \right)^{\alpha/c^2}}
                        {\left(1+\frac{r_G}{\lambda} \right)^\alpha}
                        \: R_0^{(G)}
$$
It is again straightforward to show that, all other parameters fixed, $r_S$ is an increasing function of $c$ for $c \geq 1$ and that $R_0^{(S)}$, all other parameters fixed, is a decreasing function of $c$ for $c \geq 1$. Thus, use of a Gamma distribution with larger variance than the generation time distribution but same mean always leads to overestimation of the exponential growth rate and underestimation of the basic reproduction number, given the true value of the other parameter, since the value $c=1$ corresponds to the true value generated by the generation time distribution.

\section{Modelling multiple exposures and estimating the incubation period}\label{app-mult}
As discussed in Section \ref{sec-multiple}, in order to estimate the parameters of the incubation time distribution and the infection probability per contact $p$, once assigned a parametric model, it should be possible to use Maximum Likelihood techniques on the relevant part of Equation (\ref{single-lik}). One might also try to find moment relations that allow estimation of $p$ and mean and variance of the incubation time distribution.

Suppose that the "contact" process is modelled as a Poisson process with constant rate $\mu$ with $t=0$ at the first contact with an infective,  that the probability of infection per contact is $p$, independently at each contact, and that the time from infection to symptoms is denoted by $T$, with mean $E(T)$ and variance $Var(T)$.

Then the index $I$ of the contact that infects will be Geom($p$), i.e.\ $P(I=k)=p(1-p)^{k-1}, k=1,\ldots$. After that, the number of contacts $S$ before symptoms appear will have a Poisson distribution with mean $\mu T$, given $T$. The number of observed contacts will be $C=I+S$, with summands independent. The relations
$$E(C)= 1/p + \mu E(T)$$
and
$$Var(C)=(1-p)/p^2 + \mu E(T) + \mu^2 Var(T)$$
will then hold.

Denote by $S$ the time from first contact to symptoms. Then $S$ is the sum of the time until infection and the incubation time. The time until infection is, given I, Gamma distributed with parameters $I-1$ and $\mu$. Thus
$$E(S)=\frac{1-p}{p} \frac{1}{\mu} + E(T)$$
and
$$Var(S)=\frac{1-p}{p} \frac{1}{\mu^2}+\frac{1-p}{p^2} \frac{1}{\mu^2}+Var(T).$$
Since $E(C)$, $Var(C)$, $E(S)$ and $Var(S)$ can be estimated from data, the four equations can be used for moment estimation of $p$, $\mu$, $E(T)$ and $Var(T)$.

Both ML-estimation and moment estimation have been implemented in a small simulation experiment, with 1000 replicates of estimation based on 500 individuals, each having a number of contacts $C$ and a time $S$ between first contact and symptoms. In the simulation, it was assumed that $p=1/2$, that the incubation period had mean 11.4 ($sd=8.1$) and that contacts happened according to a Poisson process with intensity 0.0725 (mean interval between contacts $= 13.8$ days). To test the stability of the estimation methods, one simulation run was performed with Gamma-distributed incubation times, assuming that the target distribution in the ML method was effectively Gamma and another run using log-normally distributed times with the same mean and variance but still using the Gamma distribution as target in the ML estimation.

\begin{table}[]
\centering
\caption{Empirical 95\% confidence intervals of parameter estimates from 1000 simulated samples of 500 individuals with true values $p=0.5$, $E(T)=11.4$ and $s.d.(T)=8.1$. Simulations from two situations, $T$ being gamma-distributed or log-normal, and ML-estimation based on gamma-distribution in both situations.}
\vskip.3cm 
\label{Tab_simul}
\begin{tabular}{|l|c|c|c|}
\hline
Model,estimator  & $p$ (=0.5)   & $E(T)$ (=11.4) & $s.d.(T)$ (=8.1)   
\\
\hline
$T\sim \Gamma$, ML            & (0.500, 0.503) & (11.41, 11.47)   & (8.11, 8.18)       
\\
$T\sim \Gamma$, Mom           & (0.503, 0.508) & (11.39, 11.63)   & (7.57, 8.22)        
\\ $T\sim Log N$, ML           & (0.486, 0.488) & (10.80, 10.85) &  (6.33, 6.39)   
\\ $T\sim Log N$, Mom           & (0.502, 0.507) & (11.36, 11.61) &  (7.52, 8.17)   
\\
\hline                                           

\end{tabular}
\end{table}

From the simulation results, shown in Table \ref{Tab_simul} one may conclude that the ML method works well if the correct distribution is assumed, but less well in case of misspecification, while the moment method seems quite stable, maybe with a hint at downward bias with the lognormally distributed data. However, further research is needed about the best moment expressions to  use and possible other approaches to this estimation problem.

\section{The simulation model for generation and serial intervals}\label{App-serial}
The representation of generation times and serial intervals in Section \ref{sec-serial} shows that $E(G)=E(S)$ but that $Var(S)=Var(G)+2Cov(s_0,\ell_0+g^*-s_0)$ (it should be noted that the distribution of the infectious period and thus of $g^*$ depends on a size-biased version of  the model infectious period (see e.g.\ Scalia Tomba et al., 2010), but that this does not affect  the present argument). Since the simulation model assumes independence between latent period $\ell_0$ and the infectious period, upon which $g^*$ depends and that $s_0=u\ell_0$, where $u\sim \text{Uniform}[0.8,1.2]$, the difference between $Var(S)$ and $Var(G)$ can be written as $2(Cov(u\ell_0,\ell_0)-Var(u\ell_0))$. Since $E(u)=1$, this reduces to $2E(\ell^2_0)(1-E(u^2))$. Since $E(\ell^2_0)=150$ and $E(u^2)=1.0133$, we should have $Var(S)-Var(G)=3.99$. Furthermore, in the model we have $G \sim \Gamma (3,1/5)$ and thus $Var(G)$ = 75. However, in the simulation results, we find the estimates $Var(G) \approx 52.4$ and $Var(S) \approx 54.6$. It should be remembered that the generation times are observed "backwards", in which case theory predicts that the observed distribution should change from $\Gamma(3,0.2)$ to $\Gamma(3,0.2387)$, leading to $Var(G)=52.7$, which is now in good agreement with simulations. It thus appears that the "backward" observation shortening also affects the observed serial intervals and the difference between observed generation time and serial interval distributions. In this case, in order to theoretically predict the observed difference in variances, one would have to calculate the effect of "backward" observation of a generation time on the marginal distribution of the corresponding latency and incubation time.

A calculation needed in Section \ref{subsec-serial} regards the coefficients of variation of $G$ and $S$, when it is assumed that the latter is larger by a factor $c$ compared to the former. Since the means of $G$ and $S$ are the same, we will have $c^2 = Var(S)/Var(G)$ and thus $c^2 = 78.99/75=1.053$ according to the above calculations. Thus $c=1.026$ in the simulation model.

\section{Estimating the growth rate}\label{est-growth}
Estimating the growth rate of the outbreak or its doubling time or other equivalent measures (under the assumption of exponential growth) is interesting per se and is useful for predictions of the future size of the outbreak if it is assumed that the current growth rate will not change. There are several possible methods but, unfortunately, it seems difficult to evaluate them theoretically on finite samples.  We have already illustrated the use of Equation (\ref{Malthus}), but it is of course possible to estimate the growth rate directly from case data. However, there seems to be a lack of systematic evaluations adapted to infectious disease spread data. A sensible approach is then to simulate the performance of the chosen estimation method in simulations as close as possible to the data generating situation. We have therefore evaluated various data-based methods in the simulations inspired by data on the 2014 Ebola outbreak in West Africa (Team WER et al., 2014). The compared methods are:\\

a) linear regression on logarithms of cumulative numbers of notified cases,

b) linear regression on logarithms of daily numbers of notified cases,

c) taking the mean of daily ratios of successive cumulative numbers,

d) estimating $exp(r)$, the daily multiplication factor, with a branching process type estimator of the form $(n(2)+\dots .+n(K))/(n(1)+\dots +n(K-1))$, where $n(i)$ is the number of cases notified day $i$, and reporting the logarithm,

e) fitting the discretized renewal equation described in Section  \ref{sec-model} to observed incidence data, using the generation time distribution estimated from backward times. This method produces an estimate of $R_0$ and not of the growth rate $r$ but can anyway extrapolate values of future incidence.\\

Using the 1000 simulated epidemic trajectories, the estimators a) - e) of the exponential growth rate $r$ and their usefulness in predicting the epidemic size 6 weeks after reaching 4500 notified cases were tested. The methods a) - d) estimate $r$ using data from the last 6 weeks before reaching level 4500, while the fifth method, based on the discretized renewal equation (Eq.\ \ref{disc-renewal}) estimates $R_0$, using regression weights derived from the estimated generation time distribution based on observed backward serial intervals, as suggested by Team WER et al.\ (2014).

In the simulation model, the true value of $r$ is 0.03870 and of $exp(r)$ is 1.03946.
It should be noted that what is then used in the prediction of the situation 6 weeks later would be the factor $exp(42r)$ (with true value = 5.07973), which is also studied, both in isolation and used as predictor in combination with the last datum.

All 4 estimators of $r$ show reasonably concentrated values around the true value 0.0387, as shown in Table \ref{r-est}:

\begin{table}[ht]
\centering
\caption{Some distributional summaries for the estimators a)-d) (see text) of the exponential increase rate based on time series of notified cases. The theoretical value to be estimated is $r=0.03870$.}
\vskip.3cm
\label{r-est}
\begin{tabular}{|l| c c c c|}
\hline
 Statistic & (a) & (b) & (c) & (d)\\
 \hline
Maximum & 0.04588 & 0.04605 & 0.04583 & 0.05168\\
Median & 0.03883 & 0.03901 & 0.03885 & 0.03904\\
Minimum & 0.03272 & 0.03125 & 0.03319 & 0.02785\\
Mean & 0.03891 & 0.03905 & 0.03892 & 0.03901\\
Std Dev & 0.00220 & 0.00230 & 0.00217 & 0.00407\\
Upper 95\% Mean & 0.03905 & 0.03919 & 0.03905 & 0.03927\\
Lower 95\% Mean & 0.03877 & 0.03890 & 0.03878 & 0.03876\\
\hline
 \end{tabular}
\end{table}
 
 However, all slightly overestimate r, since the 95\% confidence intervals don't contain the true value, although by less than 1\%, on average.
If one considers the prediction factor $exp(42r)$, on average all four estimators again overestimate a little. Estimator (a) is the best, with mean 5.14729, but the $sd$ of the distribution is now 0.48000 and a 95\% prediction interval ranges from 4.2914 to 6.1641, which means $[-16\%,+21\%]$ relative to the true value.

However, if one implements the prediction by multiplying the last cumulative datum by the estimate of $exp(42r)$ and divides the result by the true cumulative datum 42 days later, there is still overestimation, by about 1\%, but the prediction interval for method (a) has shrunk a little, to $[-13\%,+18\%]$. There is no big difference between methods, but (a) seems to have a slight advantage.

Finally, we study the performance of the renewal equation (\ref{disc-renewal}) which is also used in Team WER et al.\ (2014). This approach is intended to allow estimation of $R_t$ (in our simulation, $R_t$ is kept constant  $=R_0$ all the time and the method is adapted accordingly) and to further allow prediction of cases 6 weeks after the last datum. There are many small details to decide when using this method. We have made the following assumptions and considerations:\\
- the time series of reported cases starts with day 1 when the first case is reported (= becomes symptomatic in our simulation) and goes on until the day the total 4500 is reached. The length of this series is thus different from the one counted from the introduction of the first infective.\\
- the method uses the serial time distribution, assumed Gamma, as estimated from data. However, this distribution has to be discretized to daily probabilities. We know from previous discussions that this distribution is a biased estimate of the true generation time distribution.\\
- assuming the auto-regressive Poisson model for the time series of daily cases, the estimator of $R_0$ can be explicitly deduced.\\
- with this estimated $R_0$, the time series can be brought forward until the desired prediction date is reached.\\
The results indicate that this method  underestimates $R_0$ but has good predictive power anyway, as follows:\\
-while the true value of $R_0$ in the simulation is 1.7, the mean of estimates obtained is 1.566 with 95\% confidence limits 1.564 and 1.568, the minimum value in 1000 simulations was 1.467 and the maximum 1.683;\\
- using the quotient predicted/true observed value 6 weeks later as indicator of predictive accuracy, the mean was 1.0040 with 95\% confidence limits 1.0006 and 1.0074, with minimum value 0.856 and maximum 1.213.\\
Thus, the predictor is almost unbiased and a 95\% prediction interval is [-10\%,+12\%] around the true value, which is slightly better than the previous $r$-based methods.

\section*{Additional references}

Phillips D.T. \& Beightler C. (1971) Procedures for generating gamma variates with non-integer parameter sets. WSC '71 Proceedings of the 5th conference on Winter simulation, p 421-427, ACM New York, NY, USA.

\end{document}